
%
%
\magnification=1200
\def\uplrarrow#1{\raise1.5ex\hbox{$\leftrightarrow$}\mkern-16.5mu #1}
\def\bx#1#2{\vcenter{\hrule \hbox{\vrule height #2in \kern #1\vrule}\hrule}}
\def\Im{\,\hbox{Im}\,}

\def\tr{\,{\hbox{tr}}\,}

\def\squiggle#1{\lower1.5ex\hbox{$\sim$}\mkern-14mu #1}

\def\narrower{\advance\leftskip by\parindent \advance\rightskip by\parindent}

\def\mbox#1#2{\vcenter{\hrule width#1in\hbox{\vrule height#2in
   \hskip#1in\vrule height#2in}\hrule width#1in}}
\def\eqsquare #1:#2:{\vcenter{\hrule width#1\hbox{\vrule height#2
   \hskip#1\vrule height#2}\hrule width#1}}
\def\inbox#1#2#3{\vcenter to #2in{\vfil\hbox to #1in{$$\hfil#3\hfil$$}\vfil}}
\def\strutdepth{\dp\strutbox}
\def\marbul{\strut\vadjust{\kern-\strutdepth\specialbul}}
\def\specialbul{\vtop to \strutdepth{
    \baselineskip\strutdepth\vss\llap{$\bullet$\qquad}\null}}
\def\Bcomma{\lower6pt\hbox{$,$}}    
\def\bcomma{\lower3pt\hbox{$,$}}    

\def\updots{\mathinner{\mskip 1mu\raise 1pt\hbox{.}
    \mskip 2mu\raise 4pt\hbox{.}\mskip 2mu
    \raise 7pt\vbox{\kern 7pt\hbox{.}}\mskip 1mu}}

\def\pmb#1{\setbox0=\hbox{#1}%
     \kern-.025em\copy0\kern-\wd0
     \kern.05em\copy0\kern-\wd0
     \kern-.025em\raise.0433em\box0}

\def\1{\;1\!\!\!\! 1\;}

\def\m@th{\mathsurround=0pt}
\def\upsquarefill{$\m@th\bracelu\leaders\vrule\hfill\braceru$}
\def\ope#1{\mathop{\vtop{\ialign{##\crcr
     $\hfil\displaystyle{#1}\hfil$\crcr\noalign{\kern3pt\nointerlineskip}
     \kern4pt\upsquarefill\kern4pt\crcr\noalign{\kern3pt}}}}\limits}

\def\lsim{\mathrel{\rlap{\lower4pt\hbox{\hskip1pt$\sim$}}
    \raise1pt\hbox{$<$}}}         
\def\gsim{\mathrel{\rlap{\lower4pt\hbox{\hskip1pt$\sim$}}
    \raise1pt\hbox{$>$}}}         

\def\ffdual{{N_f\over 8 \pi^2} g^2 \tr
\epsilon^{\mu\nu\rho\sigma}F_{\mu\nu}F_{\rho\sigma}}
\vsize=7.5in
\hsize=5in
\pageno=0
\tolerance=10000
\hfuzz=5pt
\baselineskip 12pt plus 2pt minus 2pt
\hfill   DFTT 62/92

\hfill October 1992
\medskip
\centerline{\bf SMALL ANGLE POLARIZATION IN HIGH ENERGY P--P SCATTERING}
\centerline{\bf THROUGH NONPERTURBATIVE CHIRAL SYMMETRY BREAKING}
\vskip 36pt\centerline{Mauro Anselmino$^{a,b}$ and Stefano
Forte$^{b}$}
\vskip 12pt
\centerline{\it Dipartimento di Fisica Teorica, Universit\`a di
Torino$^{a}$}
\centerline{\it and }
\centerline{\it I.N.F.N., Sezione di Torino$^{b}$}
\centerline{\it via P.~Giuria 1, I-10125 Torino, Italy}
\vskip 1.4in
{\narrower\baselineskip 10pt
\centerline{\bf ABSTRACT}
\vskip 10pt
\noindent We show that a large anomalous contribution due to nonperturbative
instanton-like gluonic field configurations to the axial charge of
the proton
implies high-energy spin effects in $p-p$ elastic scattering.
This is the same mechanism which is responsible for anomalous baryon
number violation at high energy in the standard model. We compute the
proton polarization due to these effects
and we show that it is proportional to the
center-of-mass
scattering angle with a
universal (energy-independent) slope of order unity.}

\vskip 1.in
\centerline{Submitted to: {\it Physical Review Letters}}
\vfill
\eject
\input harvmac
\baselineskip 24pt plus 4pt minus 4pt
There are many indications, both experimental and
theoretical,\nref\spin{See e.g. E.~Leader, in  ``Spin and Polarization
Dynamics  in Nuclear and Particle Physics'', A.~O.~Barut, Y.~Onell and A.
{}~Penzo, Eds.
(World Scientific, Singapore, 1990)}$^{(\xref\spin)}$
that
perturbative techniques may be insufficient to account for spin effects, even
at high energy. It has been suggested\nref\mea{S.~Forte, {\it Phys. Lett.}
{\bf B224}, 189 (1989); {\it Nucl. Phys.}
{\bf B331}, 1 (1990)}
\nref\meb{S.~Forte and E.~V.~Shuryak, {\it Nucl. Phys.}
{\bf B357}, 153 (1991)}$^{
(\xref\mea,\xref\meb)}$ that the so-called ``proton spin''
problem\nref\spinrev{For a review, see G.~Altarelli, in
``Proceedings of the 1989 Erice school'' (Plenum, New York
(1990). See also R.~L.~Jaffe and A.~Manohar, {\it Nucl. Phys.}, {\bf B337},
509 (1990)}$^{
(\xref\spinrev)}$
may be such an instance.
Namely,   the observed smallness of the
axial charge of the proton may be due to a cancellation between the axial
charge of quarks, which are expected to carry most of the proton's helicity,
and a chirality-violating   contribution triggered by
the axial anomaly through the nonperturbative gluon configurations required to
solve the U(1) problem.\nref\rom\rom{See
R.~Jackiw, in S.~B.~Treiman, R.~Jackiw, B.~Zumino and E.~Witten,
``Current Algebra and Anomalies'' (World Scientific, Singapore,
1985)}\nref\uoner{See R.~Jackiw, {\it Rev. Mod. Phys.}
{\bf 49}, 681
 (1977)}\nref\uonet{G.~'t~Hooft, {\it Phys. Rep.} {\bf 142}, 357
(1986)}$^{(\xref\rom - \xref\uonet)}$
This implies that, even at high energy, chiral symmetry can be violated in
the strong interactions, and entails the possibility of unusual polarization
effects.

It is the purpose of this paper to describe one such effect. We shall compute
the effective pseudoscalar (helicity-flipping)
nucleon-nucleon-$n$ gluon interaction which is induced by
the coupling of the nucleon to instanton vacuum configurations,
analogously to  what is done in treatments of baryon number violation in the
weak interactions.\nref\matt{A.~Ringwald, {\it Nucl. Phys} {\bf B330}, 1
(1990); for a review see
M.~Mattis, {\it Phys. Rep.}, {\bf B214}, 159 (1992)}$^{(\xref\matt)}$
 We
shall fix its magnitude by assuming
it to account$^{(\xref\mea, \xref\meb)}$  for the discrepancy between
the experimental and the quark model
value of the nucleon axial charge.
We shall show that interference between this interaction and
that which dominates elastic scattering leads to a
nonvanishing value of the polarization\nref\bsl
\bsl{See e.g. C.~Bourrely, J.~Soffer and E.~Leader, Phys. Rep. {\bf 59},
95 (1980)}
$P(t)$~$^{({\xref\bsl})}$ at small scattering angles.

We shall compute $P(t)$ for $t\to0$, and  show
that it satisfies a scaling law, in that
its forward limit depends only on the center of mass scattering angle
$\theta$,
and not on the energy. Indeed, we shall show that $P(t)$
saturates the kinematical bound which constrains it to vanish in the
forward direction$^{(\xref\bsl)}$, i.e.,
$P(t){\mathop \sim\limits_{t\to 0}}
\sin{\theta\over 2}$; we shall compute the slope of the $\sin {\theta\over 2}$
dependence and show it to be of order unity and energy
independent. We shall see that this is consistent with currently available
data.

Let us first briefly recall the import and meaning of the ``proton
spin'' problem.$^{(\xref
\spinrev)}$ Polarized deep-inelastic scattering
experiments provide a
measurement of
the  matrix element of the isosinglet axial current in a nucleon (with
momentum $p$
and helicity $\lambda$)
in the limit of vanishing momentum transfer $q$.
Because of the absence of a singlet pseudoscalar
Goldstone boson, this equals
(in the helicity-nonflip channel)
\eqn\mela
{\langle p,\lambda|j^\mu_5|p,\lambda\rangle=
\lim_{q\to0}G_A(q^2) s^\mu(p, \lambda)}
where $s^\mu(p, \lambda)$ is the spin four-vector associated to the given
momentum and helicity.

The quark-parton model expectation
that quarks carry most of the nucleon's helicity leads to a value
$G_A(0)\sim 0.6$,  while the experimental value
is compatible with
$G_A(0)\sim 0$.$^{(\xref
\spinrev)}$ In QCD, however, the chiral symmetry is broken by quantum
effects, hence, the singlet axial current is not conserved
and its matrix elements may differ from the
quark expectation due to the interaction. This can be seen
explicitly$^{(\xref\mea)}$ by
noticing that  the charge operator $Q_5=\int\! d^3 x\, j^0_5$ is the sum of
the canonical
fermion helicity
operator $Q_5^q$, plus a gluonic operator $Q_5^g$
which may {\it in principle} provide a large
contribution to $G_A(0)$.

An {\it explicit} mechanism which produces a
value of the matrix elements of $Q_5^g$ large and anticorrelated to
that of $Q_5^q$ has been suggested in Ref.~(\xref\meb). This is based on the
fact that the nonperturbative vacuum structure of QCD induces an effective
helicity-flipping quark-quark
interaction.\nref\effact{G.~'t~Hooft, {\it Phys. Rev.} {\bf D14},
3432 (1976)}\nref\shif{M.~A.~Shifman, A.~I.~Vainshtein and V.~I.~Zakharov, {\it
Nucl. Phys.} {\bf B163}, 46 (1980)}\nref\shu{See E.~V.~Shuryak, {\it Phys.
Rep.} {\bf 115}, 151 (1984)
}$^{(\xref\uoner,\xref\uonet,\xref\effact - \xref\shu)}$
That is, the vacuum can be
approximated$^{(\xref\uoner,\xref\uonet)}$ by a semiclassical
superposition of gauge
fields which tunnel between vacua connected by topologically nontrivial
gauge transformations (instantons).
These generate an effective fermion-fermion
interaction$^{(\xref\effact)}$
that leads to processes where $2N_f$ units of
chirality are created. More generally, since the vacuum is not an eigenstate
of chirality, $2 n$ units of chirality may be created, while the remaining
$2(N_f-n)$ go into the vacuum mean-fields$^{(\xref\uonet)}$.
This may provide a contribution to
the axial form factor $G_A(0)$.

Whereas such contribution may be
computed exactly only in simplified models, as that of Ref.(\xref\meb),
let us investigate the consequence of {\it
assuming} that such a contribution is sizable. Due to the anomaly
equation$^{(\xref\rom)}$ satisfied by
$\partial_\mu j^\mu_5$,
Eq.\mela\ implies
\eqn
\flip
{\eqalign{\lim_{q\to 0}
iG_A(q^2)q_\mu \bar u_{\lambda^\prime}(&p^\prime)
\gamma^\mu\gamma_5u_{\lambda}(p)\cr
&=\langle p,\lambda^\prime|
\left(\ffdual + \sum_{\rm flavors}
2 i m_i \bar \psi_i\gamma_5 \psi_i\right)|p,\lambda\rangle.\cr}}
The  anomaly
density on the r.h.s. of Eq.\flip\ is proportional
to the instanton density $Q(x)$: $\ffdual(x)=-2Q(x)$.
Hence, the first term on the
r.h.s. of Eq.\flip\ receives a direct contribution from the
instanton-nucleon-nucleon coupling; indeed,
in the semiclassical approximation we may view the first term on the r.h.s. of
Eq.\flip\ as the matrix element for an instanton-nucleon-nucleon transition.
If we were able to single out the instanton contribution to the r.h.s.
of Eq.\flip, then we could view Eq.\flip\ as an expression for the effective
instanton-nucleon-nucleon coupling, which has the form
\eqn\effcoup
{\langle p,\lambda^\prime|Q^{\rm Inst}(x)|p,\lambda\rangle=-
\lim_{q\to 0}
iM_NG_A^{\rm Inst}(q^2)\bar u_{\lambda^\prime}(p^\prime)\gamma_5
u_{\lambda}(p),}
where $M_N$ is the nucleon mass. Notice that the induced
effective coupling is purely helicity-flipping
(i.e. it is nonzero only if $\lambda=-\lambda^\prime)$
and chirality-flipping, even
when $M_N\not=0$ and $q\not=0$.
The energy dependence of the
forward axial form factor $G_A^{\rm Inst}(0)$ which fixes the strength of
the effective coupling \effcoup\ is entirely fixed by the fact that,
due to the topological quantization condition on the instanton
density,\nref\polo{S.~Forte, {\it Acta Phys. Pol.} {\bf B22}, 1065 (1991).
In parton language, the instanton contribution should thus be identified with
the ``quark'' component of the proton spin, which (as discussed by
Altarelli, Ref.[\xref\spinrev)] is defined as the scale
invariant eigenstate of the Altarelli-Parisi equation satisfied by
the first moments of the polarized quark and gluon distributions.}
the instanton contribution to the axial charge \mela\ is scale
independent.$^{(\xref\polo)}$ It follows that $G_A^{\rm Inst}(0)$ is a
universal,
energy-independent coupling.

However, the r.h.s of Eq.\flip\
also receives contributions from non-instantonic field
configurations,\nref\vene{Notice that the naive identification
of the two terms
on the r.h.s. of Eq.\flip\ with a gluon and a quark
contribution, respectively, is incorrect. See
G.~M.~Shore
and G.~Veneziano {\it Phys. Lett.} {\bf B244}, 75 (1990);
{\it Nucl. Phys} {\bf B381}, 3 (1992)}$^{(\xref\vene)}$
thus, there is no way to disentangle the instanton contribution to Eq.\flip,
except by looking at other processes.
As is well known,$^{(\xref\matt)}$ in the semiclassical limit
the non-vanishing chirality-flipping amplitude for
instanton-quark-quark processes
also implies the non-vanishing of processes with the same number
of quarks and $n$ extra gluons.
That is, if $G_A^{\rm Inst}(0)\not=0$, then
Eq.\effcoup\ implies the
existence of the effective nucleon-nucleon- $n$ gluon
coupling:
\eqn\effglu
{\langle p,\lambda^\prime;\;g_1,\dots,g_n|p,\lambda\rangle=
\lim_{q\to 0}
iM_NG_A^{\rm Inst}(q^2)\bar u_{\lambda^\prime}(p)\gamma_5
u_{\lambda}(p)\prod_{i=1}^n\left[\left({2\over g_s}\right)\eta^{a_i}_{\mu_i
\nu_i}
k_i^{\mu_i}\rho^2\right],}
where $g_s$ is the strong coupling, $\eta^{a_i}_{\mu_i\nu_i}$ are the 't~Hooft
symbols,$^{(\xref\effact)}$ $a_i$, $k_i^{\mu_i}$, and $\nu_i$ are respectively
the
color\nref\ncol{Eq.\effglu\
is written for simplicity
in the case of  gauge group SU(2); in the physically relevant case
of QCD this must be embedded into SU(3),
as discussed in Ref.(\xref\shif).}$^{(\xref\ncol)}$ indices,
four-momenta,
and Lorentz indices of
the $i$-th gluon, and the external gluon propagators
have been amputated but there are no wave functions on the gluon legs.
The characteristic
parameter of the background field $\rho$  (instanton
radius)$^{(\xref\shu)}$
should be integrated over; we shall replace it with its
mean value $\rho_0$, determined phenomenologically.$^{(\xref\shu)}$

A nonzero value of $G_A^{\rm Inst}(0)$ can be tested by looking at
processes induced by the effective coupling \effglu\ with a distinct signature.
The peculiar helicity
structure of the nucleon line in Eq.\effglu\ suggests
to look at single-spin effects. These are generated
by interference between the (elastic) helicity-nonflip amplitudes
and the  amplitude where
only one of the nucleons' helicities is flipped (and are accordingly quite hard
to reproduce using standard QCD techniques).$^{(\xref\spin,\xref\bsl)}$

We shall model the elastic process
with Landshoff's nonperturbative Pomeron,\nref\lanpap{P.~V.~Landshoff and
J.~C.~Polkinghorne, {\it Nucl. Phys.} {\bf B32}, 541 (1971)\semi
G.~A.~Jaroskiewicz and P.~V.~Landshoff, {\it Phys. Rev.} {\bf D10}, 170
(1974).}
\nref\lannach{P.~V.~Landshoff and O.~Nachtmann, {\it Z.~Phys.}
{\bf C35}, 405 (1987)}$^{(\xref\lanpap,
\xref\lannach)}$
where $p$---$p$ scattering at small angles is given
by  a single diagram  which describes $t$-channel exchange of a
coherent, color-singlet state of gluons (Pomeron).
This model is phenomenologically
quite successful.  A single-helicity-flip amplitude
(Fig.~1) is
obtained by assuming that the gluons emitted by the effective instanton-induced
interaction \effglu\ ``hadronize'' into the Pomeron, which
is essentially
a two-gluon state.\nref\ingel{G.~Ingelman and P.~E.~Schlein, {\it Phys. Lett.}
{\bf B152}, 256 (1985)}$^{(\xref\lannach,\xref\ingel)}$
Typical gluon multiplicities associated to semiclassical instanton-induced
gluon emission are expected to be
$\lsim 1/\alpha_s$;$^{(\xref\matt)}$ hence,
we assume that at small momentum transfer the color-singlet
component of the effective coupling \effglu, in the case $n=2$, ``hadronizes''
into the Pomeron state with a probability of order unity. The same
assumption for the inverse process --- namely that a Pomeron emitted
by a proton in the $t$-channel at small $t$
fragments with unit probability into a two-gluon
state  --- leads to
predictions in good agreement with experiment for soft diffractive
nucleon-nucleon
scattering.\nref\diff{A.~Donnachie and P.~V.
{}~Landshoff, {\it Nucl. Phys.} {\bf B244}, 322 (1984) and
{\bf B303}, 634 (1988)\semi A.~Sch\"afer,
O.~Nachtmann, and  R.~Sch\"opf, {\it Phys. Lett.} {\bf B249}, 331
(1990)}$^{(\xref\ingel,\xref\diff)}$.

We compute the amplitude displayed in Fig.~1
by projecting the color-singlet component of a two-gluon state
generated via Eq.\effglu\
on the Pomeron with a
hadronization constant $C_H$, of order unity.
We get
\eqn\aflip
{\eqalign{M_{++;+-}=& iM_nG_A^{\rm Inst}(t)\bar u_{+}(p_1^\prime)\gamma_5
u_{-}(p_1)\left({2\over g_s}\right)^2 \cr
&\qquad\times C_H
\langle f^\mu|k^2\rangle\rho^4_0  (3\beta) F_1(t) D^2(t)
\bar u_{+}(p_2^\prime)\gamma_\mu
u_{+}(p_2)\cr}}
where $f^\mu$ is a Pomeron wave function.
The parameter $\beta=1.8\quad{\rm GeV}^{-1}$,
the elastic form factor $F_1(t)$, and the
Pomeron propagator $D(t)$
are characteristic of the helicity-nonflip
amplitude which dominates small-angle elastic scattering:$^{(\xref\lanpap,
\xref\diff)}$
\eqn\anflip
{M_{++;++}= \bar u_{+}(p_1^\prime)\gamma^\mu
u_{+}(p_1)\left[(3\beta) D(t) F_1(t)\right]^2
\bar u_{+}(p_2^\prime)\gamma_\mu
u_{+}(p_2).}
This is known
to be phenomenologically rather accurate when $-t \lsim .3\; {\rm
GeV}^2$.

The polarization is:$^{(\xref\bsl)}$
\eqn\pol
{P(\theta)\equiv{ \sigma_{\uparrow}-\sigma_{\downarrow}\over
\sigma_{\uparrow}+\sigma_{\downarrow}}=-2{\Im
\left[\phi_5\left(\phi_1+\phi_2+\phi_3-\phi_4\right)
\right]
\over|\phi_1|^2+
|\phi_2|^2+|\phi_3|^2+|\phi_4|^2+4|\phi_5|^2},}
where, assuming that at small $t$ and large enough $s$ the helicity
flipping processes are controlled by the nonperturbative vertex \effglu,
the five independent amplitudes$^{(\xref\bsl)}$ $\phi_i$ are either
expressed in terms of Eq.s~\aflip,\anflip, or else they are negligible at small
$t$:
\eqn\phis
{\eqalign{\phi_1=\phi_3&\approx M_{++;++}\cr
\phi_2\approx\phi_4&\approx M_{++;--}\propto \sin^2{\theta\over 2}\cr
\phi_5&\approx {1\over 2}\left(M_{++;+-}-M_{++;-+}\right)=M_{++;+-}.
\cr}}
Here $\theta$ is the center-of-mass scattering angle, and
we neglected the
contribution where the final state particles are exchanged.
This obtains
\eqn\res
{P(\theta)=\sin{\theta\over2} {4\sqrt{2}\over 3}G_A^{\rm Inst}(-t)
{M_N \rho_0^4 \langle k^2\rangle\over \beta F_1(t)
g_s^2}+O\left(\sin^2{\theta\over2}\right), }
where $\langle k^2 \rangle$ is the average square momentum of the gluons
in the Pomeron.

The form factor $F_1(t)$ is
constant to good approximation if $-t \lsim 0.1 {\rm GeV}^2$;$^{(\xref\lanpap)
}$ the detailed $t$
dependence of the form factor
$G_A^{\rm Inst}(-t)$ Eq.\effcoup, instead, is unknown. If we assume it to be
controlled by the instanton density, as the model computation of
Ref.(\xref\meb)
suggests, then its slope in $\sqrt{-t}$ is of the order of $V_4 n$ where $n$
is the instanton density, and the proton four-volume is roughly
$V_4\approx {1 {\rm Fm}^3\over\sqrt{ -t}}$. This gives a negligible
variation in the
$t$ range we are interested in, for reasonable values of the instanton density
$n \sim 1\;{\rm GeV}^4$.$^{(\xref\shu)}$ Notice that the slope
of the full axial form
factor $G_A(-t)$ Eq.\mela\ is
expected from current algebra to be about the same as
that of $F_1(t)$ Eq.\anflip.
Thus, for small enough $|t|$ the polarization is
\eqn\scal
{ P(\theta)=P_0\sin{\theta\over 2}}
where the slope parameter $P_0$ is a universal constant, given by
\eqn\pzero
{P_0={4\sqrt{2}\over 3} {G_A^{\rm Inst}(0) M_N \rho^4_0
\langle k^2\rangle\over
\beta g_s^2}.}
This is our main result.

It should be noticed that whereas the magnitude of
the polarization follows from our computation, its sign is not well determined:
the relative phase between the two amplitudes \aflip,\anflip\ which
enter the expression of the polarization \pol\ through Eq.\phis\ is fixed by
hermiticity and TCP invariance, but their relative sign cannot be established
with certainty in our model.
We can get a feeling for the size of $P_0$ by
taking the instanton radius to be  $\rho_0\approx
{1\over 2}$ Fm$^{(\xref\shu)}$, the strong coupling $g_s\sim 2$,
and estimating the average virtuality of gluons in
the Pomeron to be\nref\lev{E.~M.~Levin
and M.~G.~Ryskin, {\it Sov.~J.~Nucl.~Phys.} {\bf 34}, 619 (1982)}$^{(\xref\lev)
}$ $\langle k^2\rangle\approx (0.4\; {\rm GeV})^2$
(which corresponds to the given value of $g_s(k^2)$). Finally, we assume
$G_A\approx 0.5$ --- that is, we assume that the instanton contribution
reduces by that amount the quark model value $G_A\approx 0.6$.
With these
values we get $P_0\approx 0.8$.

Because angular momentum conservation forces$^{(\xref\bsl)}$ the
polarization to vanish at least as
$P(t){\mathop \sim\limits_{t\to 0}}
\sin{\theta\over2}$,
Eq.\scal\ shows that we predict that the kinematical bound is saturated.
Furthermore, the coefficient of the $\theta$ slope as $\theta\to 0$
is a universal energy-independent constant.
This is to be contrasted with the
behavior that one would predict naively on the basis of the expectation that
chiral symmetry be restored at high energy (thereby forbidding the helicity
flipping amplitudes like $M_{++,+-}$), namely, that the slope decreases
as $M_N\over\sqrt{s}$ (or some power of it).

Eq.\scal\ is valid for  small enough
$|t|$ and large enough $s$.
Effects which spoil its  validity
are:
The $t$-dependence of the form factors, which sets in when $-t \gsim 0.2\;
{\rm GeV}^2$; The breakdown of the simple-Pomeron exchange
mechanism for elastic scattering, which occurs at somewhat larger values
$-t \gsim 0.3\; {\rm GeV}^2$; The set-in of different
helicity-flipping mechanisms (not based upon the anomaly) which are suppressed
by the restoration of chiral symmetry, i.e. by powers of
${M_N\over \sqrt{s}}$, and
are thus non-negligible only for $s\lsim 10\; {\rm GeV}^2$;
The need to retain higher order terms in $\sin{\theta
\over2}$, required above
${-t\over s}\sim 0.1$. The exchange contributions to Eq.\phis\ is
negligible within these bounds.
In sum, the range of
applicability of Eq.\scal\ is $s\gsim 10\;{\rm GeV}^2$, $-t \lsim 0.1\; {\rm
GeV}^2$.

Let us review
the reliability and robustness of our results.
Our main assumption is that the instanton-induced helicity-flipping
coupling \effcoup\
suggested in Refs.(\xref
\mea,\xref\meb) may be treated semiclassically in an instanton
model in order to relate it to the amplitude with gluon emission \effglu.
Whereas the details of the computation leading to Eq.\scal\
depend on this assumption, the main feature of the result \scal, namely the
scaling law which it displays, does not. Indeed, a nonvanishing amplitude
$\phi_5$ Eq.\phis\ which
does not decrease as $M/\sqrt{s}$ is possible only if chiral symmetry
is broken at
arbitrarily high energy, thus it is a signal of an effect induced
through the anomaly by nonperturbative field configurations.
The relative phase which is required in order to get nonvanishing
polarization according to Eq.\pol\ is necessarily present,
because the anomaly equation implies that such a coupling must be
pseudoscalar (Eq.\flip), and this has the required phase by hermiticity and TCP
invariance.  Thus, whereas the instanton provides a well-defined model to
perform computations, a scaling law for the polarization Eq.\scal\ is the
signal
that  the much more general$^{(\xref\uoner)}$ phenomenon of
nonperturbative vacuum tunneling is at work.
On the contrary, the numerical value of the coefficient $P_0$ should
be taken as an order-of-magnitude estimate, since it is sensitive
to poorly known theoretical parameters such
as $\rho$ and $\langle k^2\rangle$.

Let us finally turn to the experimental situation. In order to test Eq.\scal,
ideally one would need to plot a few values of the polarization for $-t\lsim
0.1\;{\rm GeV}^2$ at various energies  $s\gsim10\; {\rm GeV}^2$, and check
that they all lie on the same
line.
A few data\nref\dataa{M.~Borghini {\it et al.} {\it Phys. Lett.} {\bf B36}, 501
(1971)}\nref\datab{D.~G.~Crabb {\it et al.} {\it Nucl. Phys.} {\bf B121},
231 (1977)}\nref\datac{S.~L.~Kramer {\it et al.} {\it Phys. Rev.} {\bf D17},
1709 (1978)} \nref\datad{A.~Gaidot {\it et al.} {\it Phys. Lett.} {\bf B61},
103 (1976)}\nref\datae{J.~H.~Snyder {\it et al.}
{\it Phys. Rev. Lett.} {\bf 41}, 781 (1978)}
are available$^{(\xref\dataa-\xref\datae)}$, despite the small values of the
scattering angle involved.
In Fig.2 we plot the slope parameter $P_0$  extracted from these data
sets
by taking the point at the lowest available value
of $t$  for each  set, and assuming that it obeys the law
Eq.\scal. Clearly,  the data do show a large value of $P_0$,
which seems  energy-independent to very good approximation. Notice that
the  data point with the lowest value $s=10$ GeV$^2$ is at the
extreme of the allowed $s$ range.
A more stringent test of our
prediction would require the availability of several values of the polarization
at different small angles. These currently are not available, perhaps also due
to lack of theoretical motivation. If experimentally confirmed, this would
provide the first direct evidence for non-perturbative QCD effects.

In conclusion, it is interesting to observe that due to the universality of
both the Pomeron and the anomalous coupling$^{(\xref\polo)}$ these effects
should be present also in different related processes, such as $p$--$\bar p$
or $\pi$--$p$ scattering, which are currently under investigation.

\bigskip
\noindent{\bf Acknowledgment:} We thank E.~Predazzi for discussions.

\vfill
\eject
\listrefs
\vfill
\eject
\centerline{\bf FIGURE CAPTIONS}
\item{[Fig.~1]} The helicity-flipping amplitude Eq.\aflip
\medskip
\item{[Fig.~2]} The value of the forward polarization slope $P_0$
[Eq.\scal] versus $s$ (in GeV$^2$). The cross indicate data from
Ref.(\xref\dataa),
the diamond   data from Ref.(\xref\datab),  the square the data from Ref.
(\xref\datac), the dash data from
Ref.(\xref\datad), and the star data from Ref.(\xref\datae).
The solid line is drawn through the mean value $P_0=0.83$.
\vfill
\eject
\bye